# Self-Organized Criticality: A Prophetic Path to Curing Cancer

J. C. Phillips

Dept. of Physics and Astronomy, Rutgers University, Piscataway, N. J., 08854

1-908-273-8218     jcphillips8@comcast.net

**Abstract**

While the concepts involved in Self-Organized Criticality have stimulated thousands of theoretical models, only recently have these models addressed problems of biological and clinical importance. Here we outline how SOC can be used to engineer hybrid viral proteins whose properties, extrapolated from those of known strains, may be sufficiently effective to cure cancer.

  Self-Organized Criticality (SOC) is a concept that has enormous intuitive appeal, which explains why it has been widely discussed, especially in biological contexts [1-3]. To extend these discussions to clinically relevant contexts, one should develop methods for engineering proteins that involve suitable general concepts. Perhaps most fundamental is the globularly compacted character of protein folds, driven by hydropathic forces. All protein chains are folded into globules by water pressure, as proteins have evolved as chains with alternating hydrophilic and hydrophobic segments. There are hundreds of thousands of strains of common viruses, whose full amino acid sequences are known. Here I analyze these sequences quantitatively, and show that their hydropathic evolution has easily recognized Darwinian (punctuated) features. These features are global in nature, and are easily identified mathematically using global methods based on thermodynamic criticality and modern mathematics (post-Euclidean differential geometry). While the common influenza virus discussed here has been rendered almost harmless by decades of vaccination programs, the hierarchical sequential decoding lessons learned here are applicable to other viruses that are emerging as powerful weapons for controlling and even curing common organ cancers.

Our first step is to realize that the protein folding problem, which has been the subject of $> 10^5$ papers, is of little or no concern here, because it has been largely solved experimentally through structural models obtained by diffraction from protein crystals. These models have revealed many details of static protein structures, which are often sufficient to explain qualitative aspects of protein functionality, yet they have left unexplained many features of protein behavior, such as substantial changes in effectiveness with 1% or fewer mutations of their amino acid (aa) side groups, or the fact that as much as 50% such mutations can occur and yet leave the fold represented by the protein's peptide chain virtually unchanged to the accessible resolution limit of 0.5A [4,5]. Backbone stability can be turned to advantage by assuming that few % mutations of wild protein strains often leave the fold unchanged, so that such strains belong to a common homology group. For flu glycoproteins this group is remarkably robust, as often > 10 aa deletions from the key transmembrane segment are required to alter fusogenic activity qualitatively [6].

For a homological group of proteins one can now take the next step, and pass directly to (amino acid sequence) – functional relations, without attempting to elucidate all the multiple functional details, such as oligomer formation. This large jump could succeed if we had access to a very accurate hydropathic scale that quantifies the large chain curvatures associated with hydrophobic aa, reflecting their tendencies to be in the globular interior, and the much smaller curvatures associated with hydrophilic aa, reflecting their tendencies to be on the globular surface. It is just here that SOC comes to our aid, in the form of a remarkable discovery [7,8] that has gone almost unnoticed. One uses Voronoi partitioning to construct polyhedra centered on each aa in turn, and from these calculates the solvent accessible (for a 2 A spherical water molecule) surface area A for each aa, averaged over a very large number of structures in the Protein Data Base.

If we average these A over complete protein structures, we in effect assume that the water-driven curvatures are independent of the protein fold. Less restrictively, assume that the aa-specific effective curvatures depend on the length $2N + 1$ of a protein segment centered on the aa, and plot A as a function of N. One then obtains [7] a stunning and very profound result: for $4 < N < 17$, $A \propto N^\delta$, which is power-law (self-similar) scaling (SOC). The exponents $\delta$ define an SOC-based hydropathicity scale $\psi(aa)$, which has turned out to be much more useful in most cases than all older $\psi$ scales, and much more accurate. Nor is that all: the lower limit of the power law range, $N = 4$, is probably related to the pitch of Pauling's $\alpha$ helix, while the upper limit is typical of the lengths of "primers" used in aa substitution (mutagenesis). The center of the range, $N = 10$, corresponds to a typical length of a membrane-spanning protein segment.

With these tools we can now proceed to analyze the mutations of flu viral strains (1945-2011). We chose these for several reasons: (1) they are mutation-prolific, and Web-accessible data bases (Uniprot, NCBI) contain extensive aa sequence data, and (2) flu proteins are very similar to Newcastle Disease Virus (NDV). NDV has been known for 50 years to select and kill cancer cells, but at present the effectiveness of the best strains is still limited to small-animal tumors. The strains cannot destroy human-size tumors before they are destroyed by antibodies [9]. Our aim is to engineer these strains so that they can work more effectively against cancer cells, while causing even less damage to healthy cells than known strains (which already have only weak, flu-like effects).

The next step is to realize that because water-protein interactions are weak and π-like (compared to protein-protein backbone interactions, which are strong and σ-like), their effects will be averaged over certain length scales W to obtain convoluted aa profiles ψ(aa, W). It is here that specific protein features enter the calculation. For flu proteins these features are either the spacing of relevant glycosidic (sugar) sites along the aa chain, or the length of sialic acid molecules to which the viral protein is bound. These spacings are known from structural studies, and as a check on the method, they also are confirmed by conservation of ψ(aa, W) profile features under migration and vaccination pressures. While the following results have been obtained using the SOC scale to identify and hierarchically quantify strain sequence differences, much of the data base searching has utilized the Web-based Basic Linear Alignment Search Tool (BLAST). BLAST is an advanced genetic similarity tool which yields interesting results for flu even without utilizing hydropathicity hierarchies [10].

Influenza virus contains two highly variable envelope glycoproteins, hemagglutinin (HA) and neuraminidase (NA). The structure and properties of HA, which is responsible for binding the virus to the cell that is being infected, change significantly when the virus is transmitted from avian or swine species to humans. Enzyme NA cleaves sialic acid groups and is required for influenza virus replication. The average HA and NA hydropathicities <ψ(aa,1)> are distinctly hydrophilic, especially HA (Table I), while NA <ψ(aa,1)> exhibits the effects of species evolution compaction. Next we optimize W to $W_{max}$ to obtain panoramas of NA and HA evolution. Details are given elsewhere [11,12], but some highlights are the following.

The simplest evolution, which exhibits opposing effects of stabilizing migration and vaccination evasion, is that of the NA variance $<<\psi(aa,17)>^2> - <<\psi(aa,17)>>^2$. The variance measures the "ball bearing" roughness of the protein's water package as it slides and/or tumbles into successive conformations. The variance decreases as the virus adapts to vaccination pressures, or increases as the viruses are compacted by migration (Fort Dix, swine flu). Because of SOC, these changes are often punctuated, as shown in the sketch of Fig. 1, or as presented and discussed in tabular form elsewhere [12].

The HA evolution is more complex, as its interaction with sialic acid (a large molecule) separates HA into blocks with coordinated responses and companion punctuations with NA. The blocks shift abruptly with each critical punctuation, as expected from SOC (often used to describe earthquakes). The one-dimensional character of these HA proteinquakes extends along the protein chain connecting non-Euclidean block edges which are contact points to sialic acid in Euclidean structures determined by diffraction. Mutations shift entire non-Euclidean blocks that exist in hydropathic coordination space. See the example in Fig. 2 for the Fort Dix outbreak, whose origin is readily identified with BLAST [12].

In conclusion, by exploiting all available modern mathematical and physical tools, as well as the very large data bases available for several viruses, one can extract very accurately features of protein functionality directly from amino acid sequences alone. Analysis similar to that presented here for the flu glycoproteins NA and HA has been applied to the Newcastle Disease Virus (NDV) glycoproteins F and HN, in order to increase the NDV oncolytic (tumor-destroying) effectiveness [12,13]. The analysis had an unexpected result: it predicts that the hybridized strains that are more effective in destroying cancer tissues may be less virulent against normal tissue. This makes massive doses by arterial injection possible [14], curing not only large tumors, but metastasized cancers as well. By far the most successful strain used in [14] contains only one new engineered NDV F amino acid mutation (out of >550 aa), leaving ample room for more fold-preserving engineering of hybridized strains.

I am grateful to Prof. M. W. Deem for critical comments.

| Protein (Species) | <MZ> | Protein (Species) | <MZ> |
|---|---|---|---|
| Adren (H) | 154.7 | Lyso (C) | 153.1 |
| Rhodop (L) | 167.1 | Lyso (H) | 154.7 |
| Rhodop (C) | 167.6 | NA | 151.5 |
| Rhodop (H) | 167.8 | HA | 148.9 |

Table I. Hydropathicities $<\psi>$ for the SOC scale [6]. Shown are lysozyme *c* (aka chicken egg white, hydroneutral), adrenergic ($\beta 1$) (membrane signaling protein that binds adrenalin), and rhodopsin (retinal membrane optical signaling protein) values for several species (Lamprey (most primitive vertebrate), Chicken, Human). In general evolution stabilizes proteins by compacting them and increasing $<\psi>$. Note that rhodopsin is exceptionally stable, as it must be to receive and process optical signals. NA is noticeably hydrophilic, and HA is even more hydrophilic (less stable).

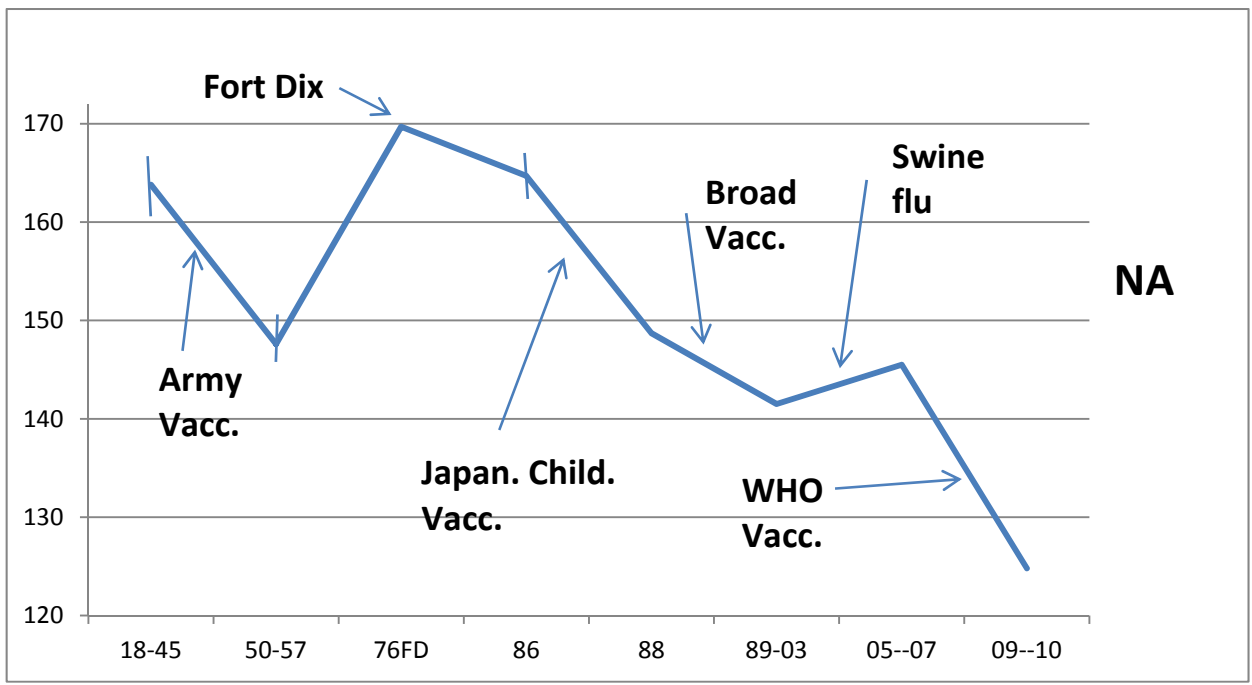

Fig. 1. The opposing effects of migration and vaccination on NA roughness (variance of $\psi(aa,17)$, MZ scale) after the first wide-spread vaccination program, begun by the US Army in 1944. Flu virulence decreases or increases in tandem with NA roughness. The build-up of swine flu from 2001 on is evident in selected urban areas with crowded immigrant neighborhoods [11]. A few early error bars are indicated, but after 1986 these become too small for this sketch.

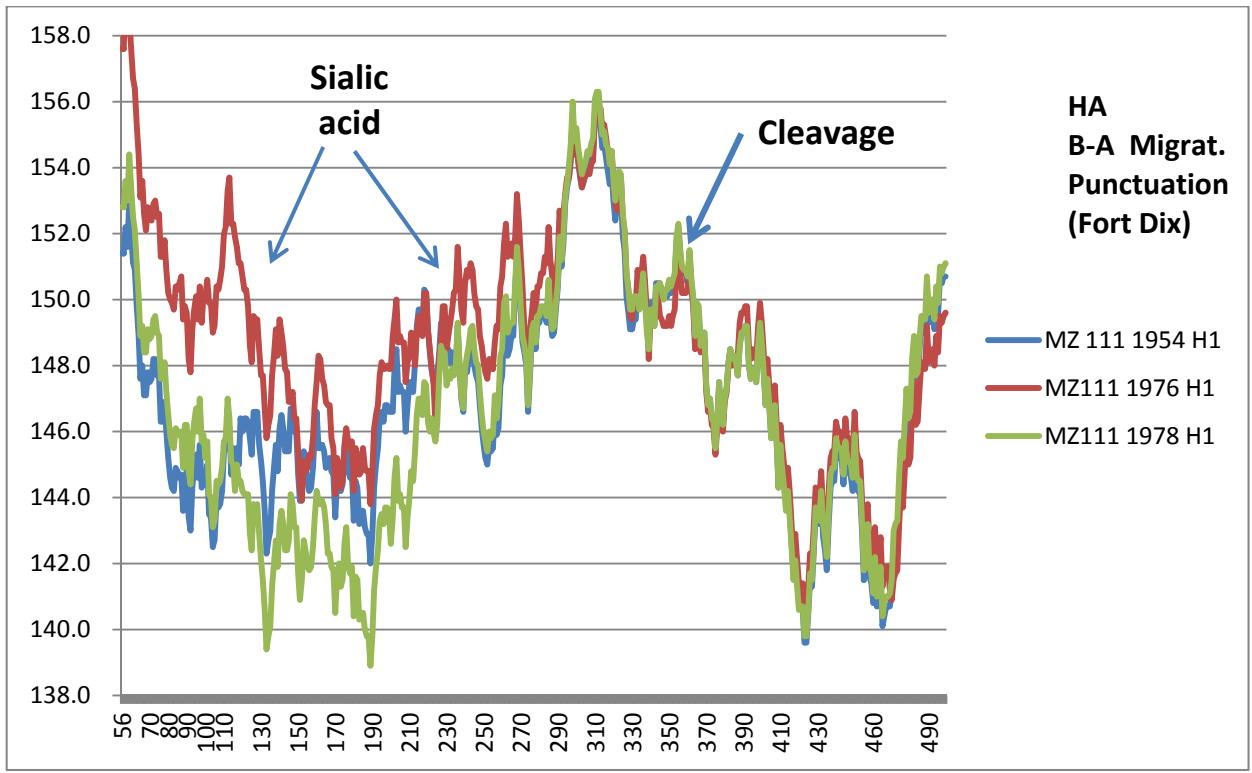

Fig. 2. Hydropathic profiles of $<\psi(aa,111)>$ for HA aa sequences before, during, and after the 1976 Fort Dix outbreak, which produced an immediate vaccination program whose effects are seen in 1978 profiles.